\documentclass[conference,letterpaper,10pt]{IEEEtran}
\usepackage{latexsym}
\usepackage{amsmath}
\usepackage{amsfonts,amssymb,bm}
\usepackage{array,cite,eqparbox,dcolumn}
\usepackage{graphicx}
\graphicspath{{figures/pdf/},{figures/png/}}
\DeclareGraphicsExtensions{.pdf,.png}
\usepackage{subfig}

\newtheorem{theorem}{Theorem}

\def\vec#1{{\bm #1}}
\def\op#1{#1}

\def\ket#1{| #1 \rangle}
\def\bra#1{\langle #1 |}
\def\ip#1#2{\langle #1 | #2 \rangle}

\def\RR{\mathbb{R}}

\def\Sgn{\operatorname{Sgn}}

\def\H{\mathcal{H}}

\def\RR{\mathbb{R}}

\def\sz{\op{\sigma}_z}
\def\eps{\epsilon}

\def\up{{\uparrow}}
\def\dn{{\downarrow}}

\def\semph#1{\textit{\textbf{#1}}}

\begin{document}
\title{Geometry and Curvature of Spin Networks}
\author{\IEEEauthorblockN{E Jonckheere\IEEEauthorrefmark{1}, 
                          S G Schirmer\IEEEauthorrefmark{2} and
                          F C Langbein\IEEEauthorrefmark{3}}
\IEEEauthorblockA{\IEEEauthorrefmark{1}
     Ming Hsieh Department of Electrical Engineering,
     University of Southern California, Los Angeles, CA 90089-2563\\
     Email: jonckhee@usc.edu}
\IEEEauthorblockA{\IEEEauthorrefmark{2} 
     Dept of Applied Maths \& Theoretical Physics, 
     Univ. of Cambridge, Cambridge, CB3 0WA, United Kingdom\\  
     Email: sgs29@cam.ac.uk}
\IEEEauthorblockA{\IEEEauthorrefmark{3} 
     School of Computer Science \& Informatics,
     Cardiff University, 5 The Parade, Cardiff, CF24 3AA, United Kingdom\\
     Email: F.C.Langbein@cs.cardiff.ac.uk}
}
\maketitle
\begin{abstract}
A measure for the maximum quantum information transfer capacity (ITC)
between nodes of a spin network is defined, and shown to induce a metric
on a space of equivalence classes of nodes for homogeneous chains with
XX and Heisenberg couplings.  The geometry and curvature of spin chains
with respect of this metric are studied and compared to the physical
network geometry.  For general networks hierarchical clustering is used
to elucidate the proximity of nodes with regard to the maximum ITC.
Finally, it is shown how minimal control can be used to overcome
intrinsic limitations and speed up information transfer.
\end{abstract}

\section{Introduction}

Networks of interacting quantum particles --- so-called spin-networks
--- are important for transferring and distributing quantum information
between different parts of a larger system such as different quantum
components on a chip~\cite{Bose-review}.  Spin chains, linear
arrangements of spins, for example, can play the role of classical wires
connecting two parts, and branched networks allow the distribution of
quantum information to different nodes. The way quantum information
propagates through spin networks, however, is quite different from
classical information flow due to quantum inference effects.  In
particular, quantum state transfer between the nodes of the network is
limited by fundamental physical principles and perfect quantum state
transfer is usually possible only in very special cases.  Quantifying
state transfer fidelities for spin networks is not easy in general, but
for certain types of networks such as spin-$\frac{1}{2}$ particles with
interactions of so-called XXZ type, for example, this problem can be
reduced to the maximum probability for a single excitation to propagate
from one node to another, which can be computed efficiently numerically,
and in some cases analytically.  Thus, for the respective spin networks,
the latter is a basic measure for the maximum information transfer
capacity (ITC) between different nodes.  For certain types of simple
networks such as homogeneous chains the maximum ITC is shown to induce a
metric on a set of equivalence classes.  We study the topology of simple
networks with regard to this metric, showing that it differs
substantially from the physical geometry.  Networks with trivial
physical geometry such as linear chains can have a surprisingly rich
geometric structure including curvature.  Analysis of the latter shows
that spin chains appear to be Gromov-hyperbolic with regard to this
maximum ITC metric but unlike classical hyperbolic networks their Gromov
boundary appears to be a single point.  For more complex networks the
maximum ITC does not induce a metric but we can use hierarchical
clustering to assess the proximity of nodes with regard to information
transfer capacity.  Again, the resulting cluster structures differ
substantially from the neighborhood relations induced by the physical
geometry of the network, showing that the latter is not very useful in
assessing the maximum ITC between nodes in a network, unlike in the
classical case.  Finally, we consider how minimal control of a single
node in the network allows us to change information flow in the network,
effectively changing the information transfer capacities (and network
``topology''), allowing us to achieve higher state transfer fidelities
as well as generally speeding up the rate of information transfer.

\section{Information Transfer Capacity}

The Hilbert space of spin networks with XXZ interactions can be
decomposed into so-called excitation subspaces $\H=\oplus_{n=0}^N \H_n$,
where $n$ is the number of excitations in the network, ranging from $0$
to $N$.  If we denote the spin basis states by $\ket{\up}$ and $\ket{\dn}$,
taking the latter to denote the ground state, then the $0$-excitation
subspace consists of a single state $\ket{0} := \ket{\dn}\otimes \cdots
\otimes\ket{\dn}$, while the one-excitation subspace consists of $N$
states $\ket{n}:=\ket{\dn}\otimes\cdots\otimes \ket{\up} \otimes \cdots
\otimes \ket{\dn}$, where the excitation $\ket{\up}$ is in the $n$th
position.  Thus transferring a quantum state
$\ket{\psi}=\cos(\theta)\ket{\dn}+e^{i\phi}\sin(\theta)\ket{\up}$ from
spin $m$ to $n$ is equivalent to transferring an excitation $\ket{\up}$
from spin $m$ to $n$:
\begin{align*}
  \ket{\psi}_m 
  &= \cos(\theta) \ket{0} + e^{i\phi}\sin(\theta)\ket{m} \\
  &\mapsto  \cos(\theta) \ket{0} + e^{i\phi}\sin(\theta)\ket{n} = \ket{\psi}_n,
\end{align*}
where we used the shorthand $\ket{\psi}_n$ to denote a product state
$\ket{\dn} \otimes \ket{\psi} \otimes \ket{\dn}$, whose $n$th factor is
$\ket{\psi}$, all others being $\ket{\dn}$.

\begin{table*}
\[\begin{array}{llll}
\hline
N & \mbox{vertices/equivalence classes}  
  & \mbox{geometry} & \mbox{Distances}\\\hline
3 & a=\{1,3\}, b=\{2\}   
  & \mbox{single edge} 
  & \overline{ab}=0.81 \\
4 & a=\{1,4\}, b=\{2,3\} 
  & \mbox{single edge} & \overline{ab}=0.32 \\
5 & a=\{1,5\}, b=\{2,4\}, c=\{3\} 
  & \mbox{triangle} 
  & 0.33 = \overline{ac}=\overline{bc} > \overline{ab} = 0.87\\
6 & a=\{1,6\}, b = \{2,5\}, c=\{3,4\} 
  & \mbox{triangle} 
  & 0.48 = \overline{ab}=\overline{cb} > \overline{ac} = 0.36\\
7 & a = \{1,7\}, b= \{2,6\}, c=\{3,5\}, d=\{4\}
  & \mbox{triangular pyramid}
  &  \overline{ad}=\overline{bc}=\overline{cd} \gg 
     \overline{ab}=\overline{bc}=\overline{ac} \\
8 & a = \{1,8\}, b= \{2,7\}, c=\{3,6\}, d=\{4,5\}
  & \mbox{triangular pyramid}
  &  \overline{ad}=\overline{bc}=\overline{cd} > 
     \overline{ab}=\overline{bc}=\overline{ac} \\
9 & a = \{1,9\}, b=\{3,7\}, c=\{4,6\}, d=\{2,8\}, e=\{5\} 
  & \mbox{4-simplex} 
  & \overline{ab}=\overline{ac}=\overline{bc}<
    \overline{ad}=\overline{bd}=\overline{cd} \ll
    \overline{ae}=\overline{be}=\overline{ce} \\
10& a = \{1,10\}, b=\{2,9\}, c=\{4,7\}, d=\{5,6\}, e=\{3,8\} 
  & \mbox{4-simplex}
  & \overline{ab}=\overline{ac}=\overline{bc} <
    \overline{ad}=\overline{bd}=\overline{cd} <
    \overline{ae}=\overline{be}=\overline{ce}\\\hline
\end{array}\]
\caption{ITC geometry of linear Heisenberg chains.}
\label{table1}
\end{table*}

\begin{figure*}
\includegraphics[width=0.49\textwidth]{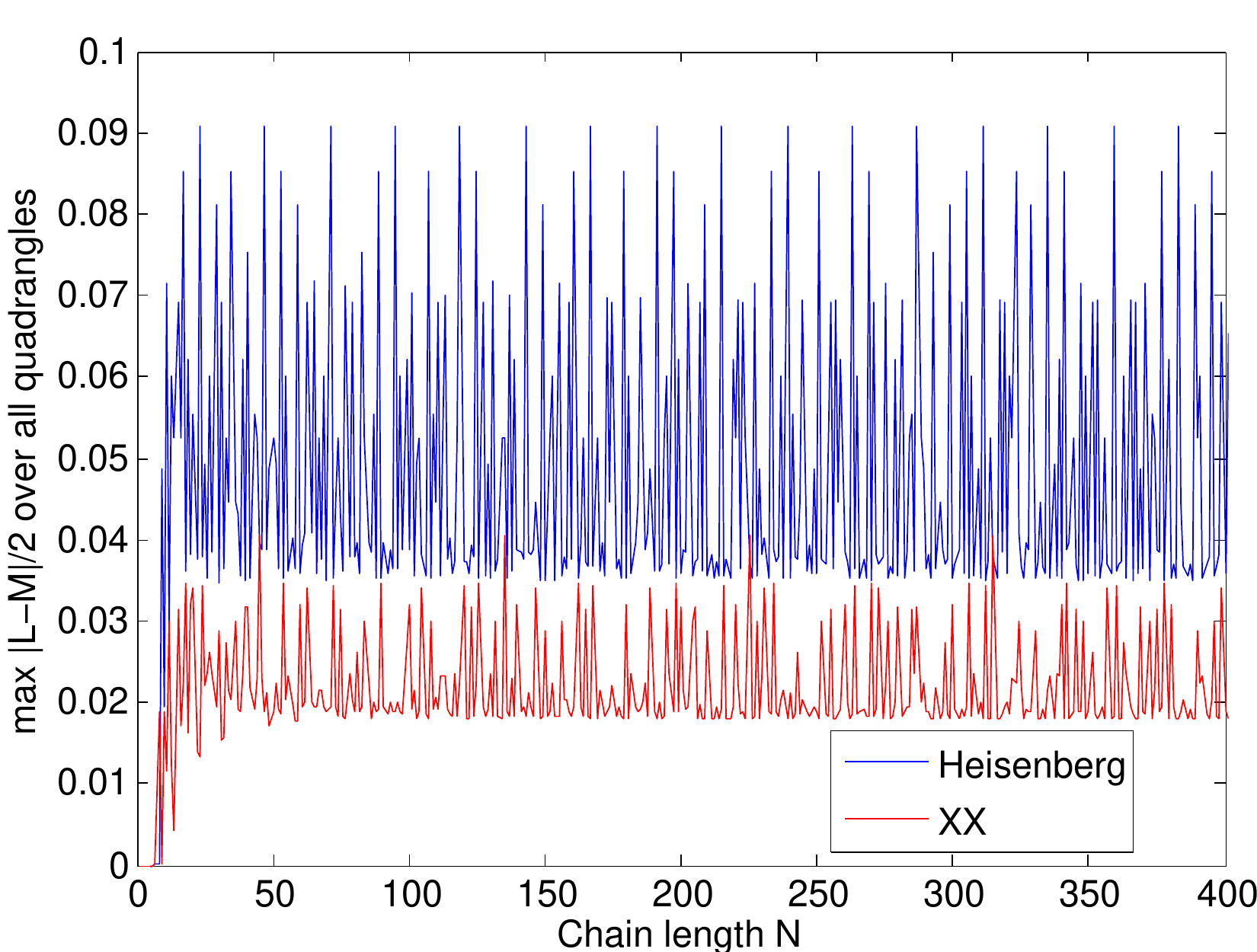}
\includegraphics[width=0.49\textwidth]{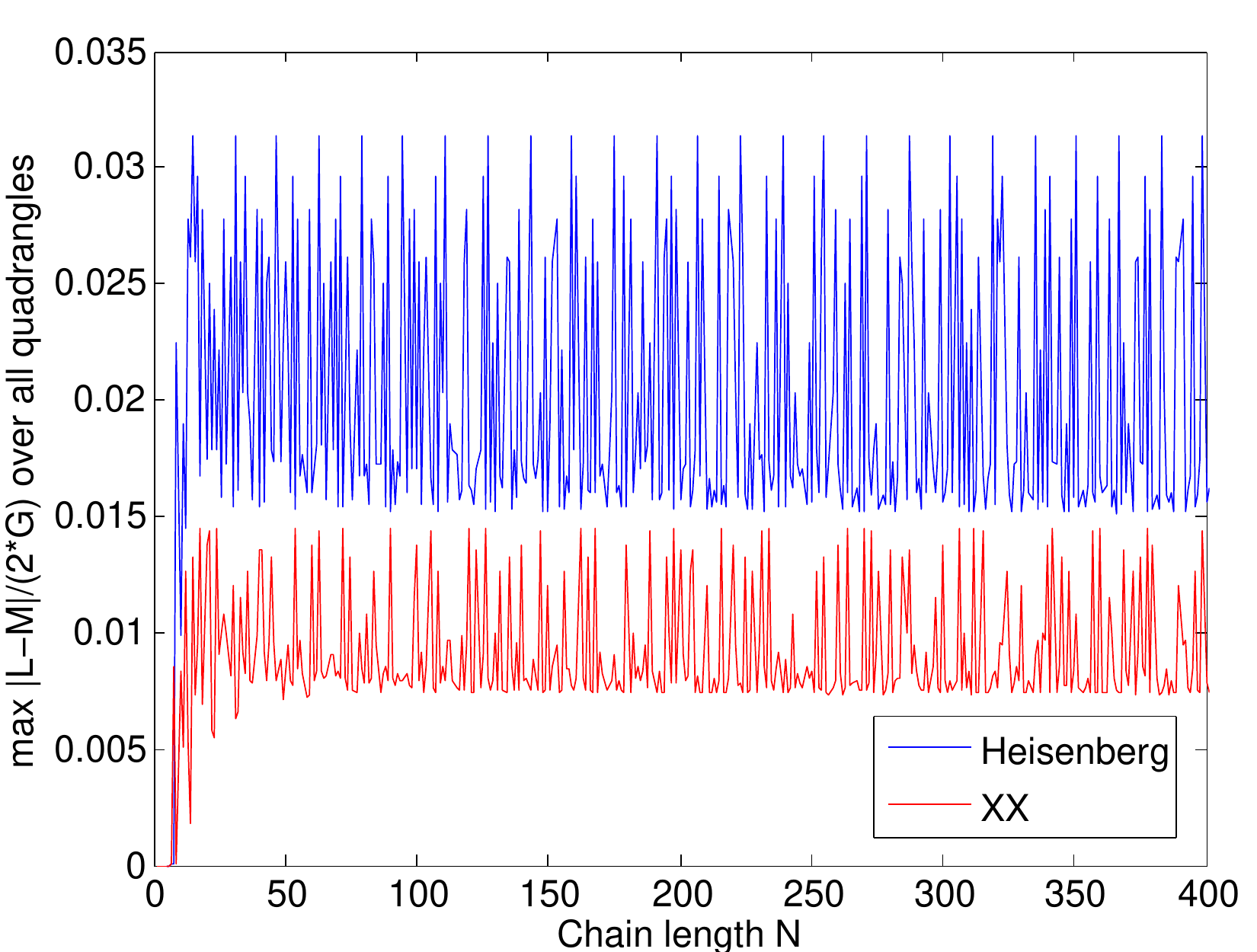} 
\caption{Chain length versus 4-point Gromov $\delta$ (left) and scaled
4-point Gromov $\delta$ (right).  Scaled-Gromov $\delta$ remains below
the upper bound singled out in~\cite{4point_Fariba}, 4-point Gromov
$\delta$ saturates at large chain length, revealing Gromov hyperbolic
property. $G=L+M+S$ where $L$, $M$ and $S$ are the pairs of opposite
diagonals of a quadrangle corresponding to the largest, medium and 
smallest length.}  \label{f:Gromov-plot}
\end{figure*}

The probability that an excitation created at site $\ket{i}$ has
propagated to site $\ket{j}$ after some time $t$ is given by
\begin{equation}
  p(\ket{i},\ket{j},t) = |\bra{j}e^{-\imath H t}\ket{i}|^2
\end{equation}
in the system of units where $\hbar=1$.  The maximum of this probability
$p(i,j)=\max_{t\ge 0} p(i,j,t)$, or a monotonic function thereof such as
\begin{equation}
  \label{eq:dist}
  d(i,j) = -\log p(i,j)
\end{equation}
gives a measure for the maximum state transfer fidelity between two
nodes in a spin network without control, quantifying the intrinsic
capacity of a spin network for quantum state transfer tasks.  It can be
shown that
\begin{align*}
 p(i,j) 
 &= \left| \bra{i} e^{-\imath H t} \ket{j} \right|
  =  \left| \sum_k \ip{i}{v_k} \ip{v_k}{j} e^{-\imath\lambda_k t}\right| \\
 &\leq \sum_k \left| \ip{i}{v_k}\ip{v_k}{j} \right|,
\end{align*}
where $H=\sum_{k=1}^N \lambda_k \ket{v_k}\bra{v_k}$ is the
eigendecomposition of the Hamiltonian operator $H$ of the network in the
first excitation subspace $\H_1$.  If the rescaled eigenvalues
$\frac{\lambda_1}{\pi}, \ldots, \frac{\lambda_N}{\pi}$ are rationally
independent~\cite[Proposition 1.4.1]{KatokHasselblatt} then it can be
shown that for any $\epsilon>0$ there exists a $t_\eps>0$ such that
$p_{\max}(i,j)-p(i,j,t_\eps))<\epsilon$, i.e., the maximum ITC is a
tight bound and attainable in the limit.  

\section{ITC Geometry \& Curvature of Chains}

\begin{figure*}
\includegraphics[width=0.49\textwidth]{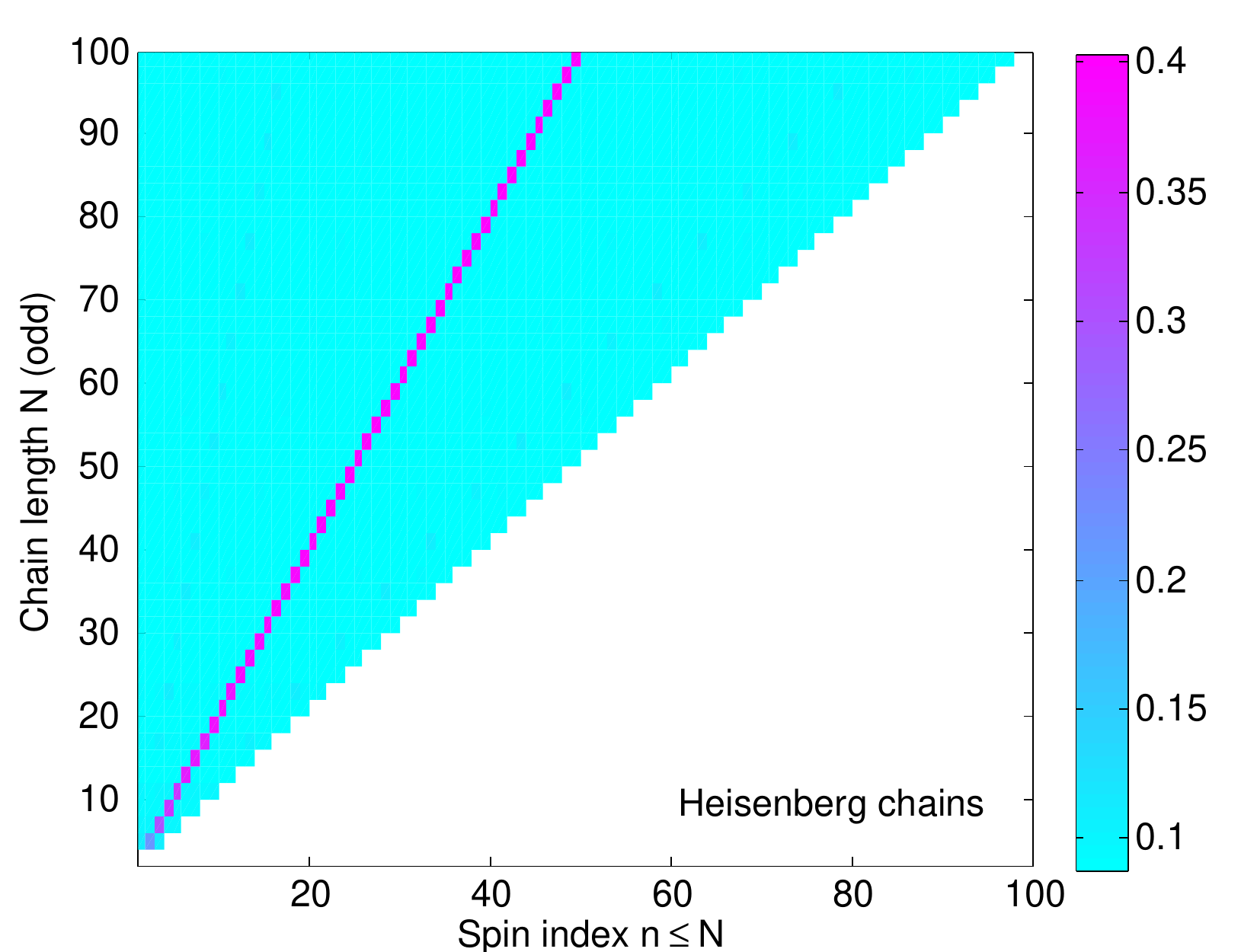}
\includegraphics[width=0.49\textwidth]{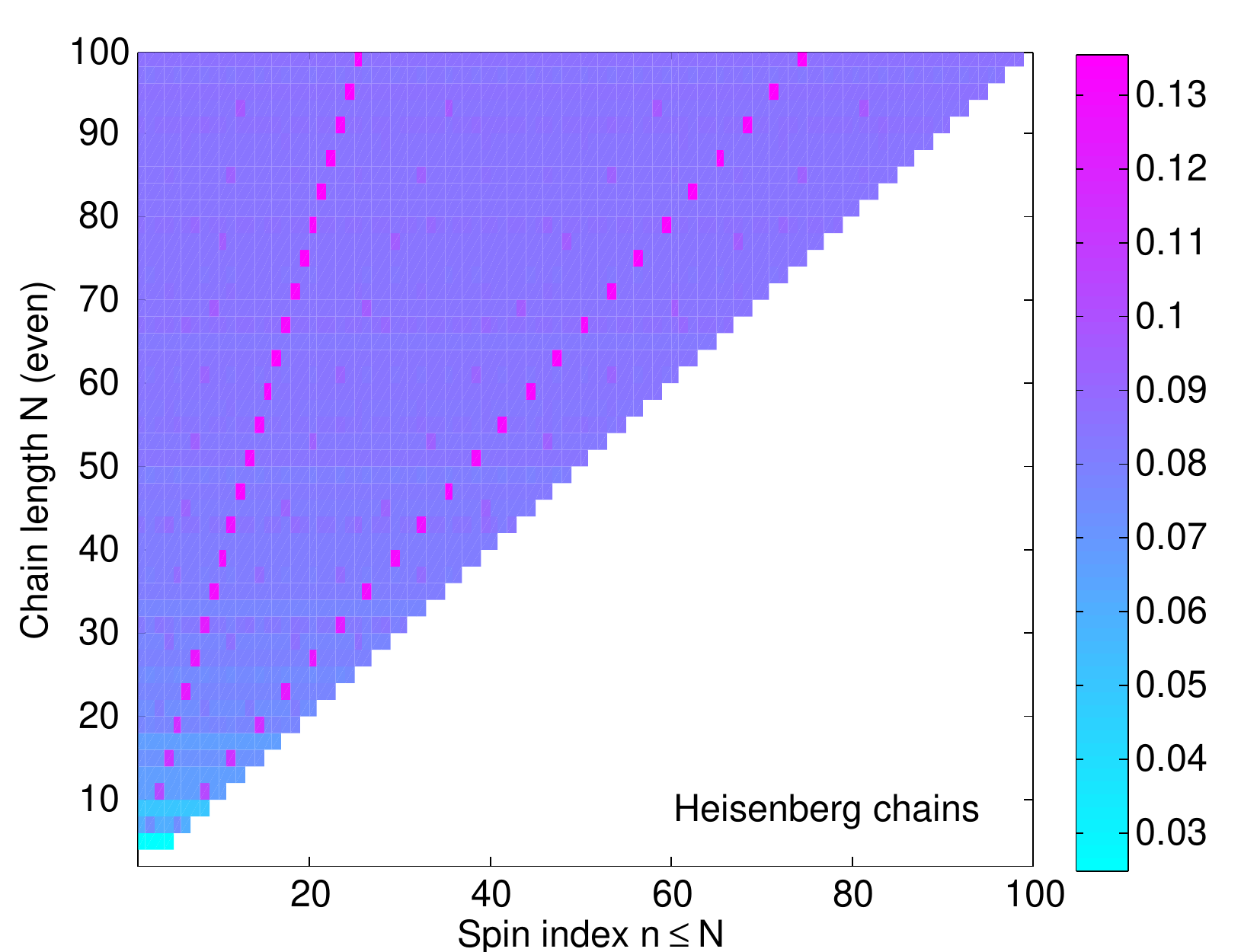}
\caption{Inertia ($\alpha=2$) versus chain length plots for Heisenberg
chains.  For $N$ odd (left) the central node is a strong anti-gravity
center.  For $N$ even (right) there are weaker anti-gravity centers
between end spins and middle (right).  The inertia is symmetric about
the central node as antipodal spins belong to the same equivalence
class.}  \label{f:InertiaPlot-H}
\end{figure*}

\begin{figure*}
\includegraphics[width=0.49\textwidth]{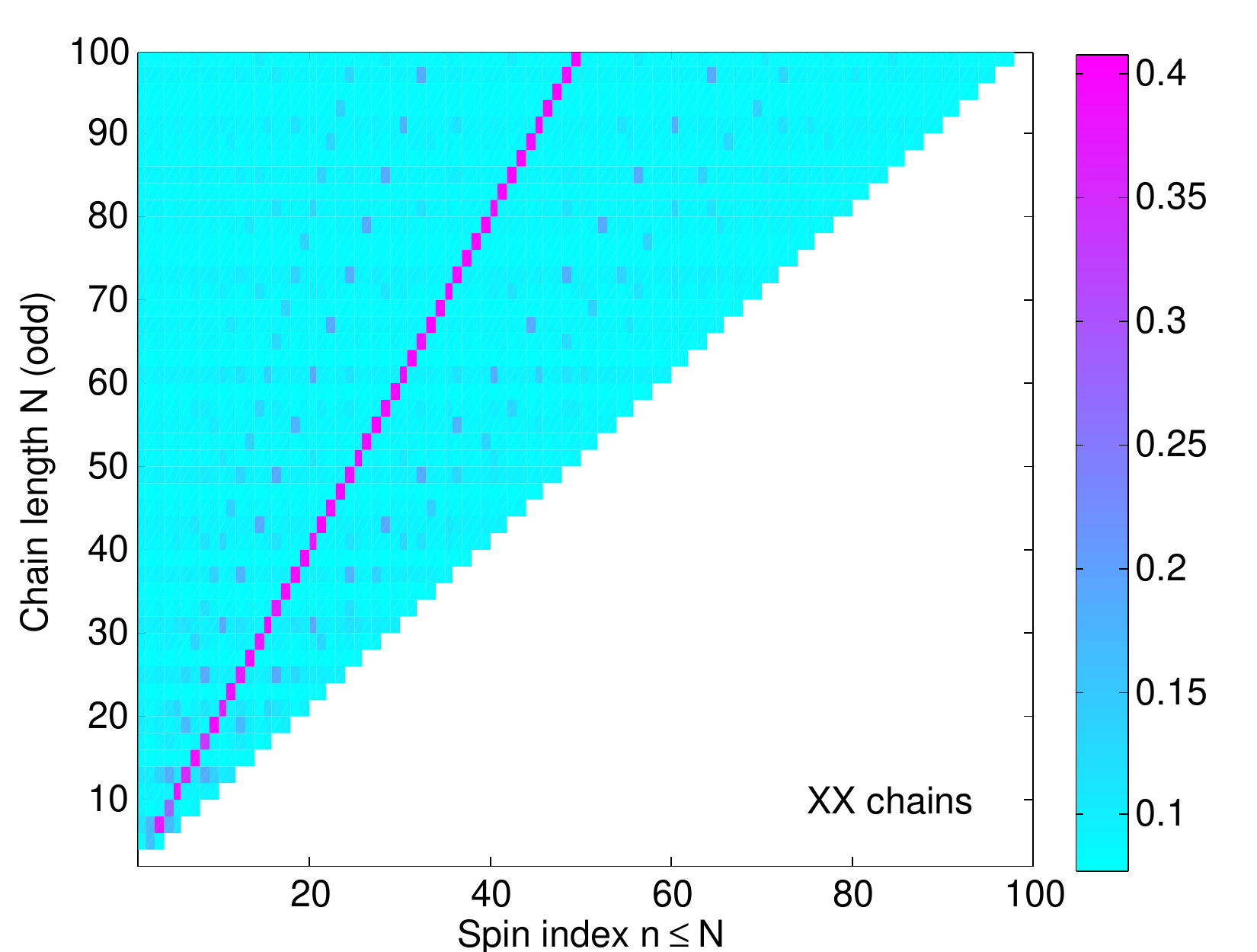}
\includegraphics[width=0.49\textwidth]{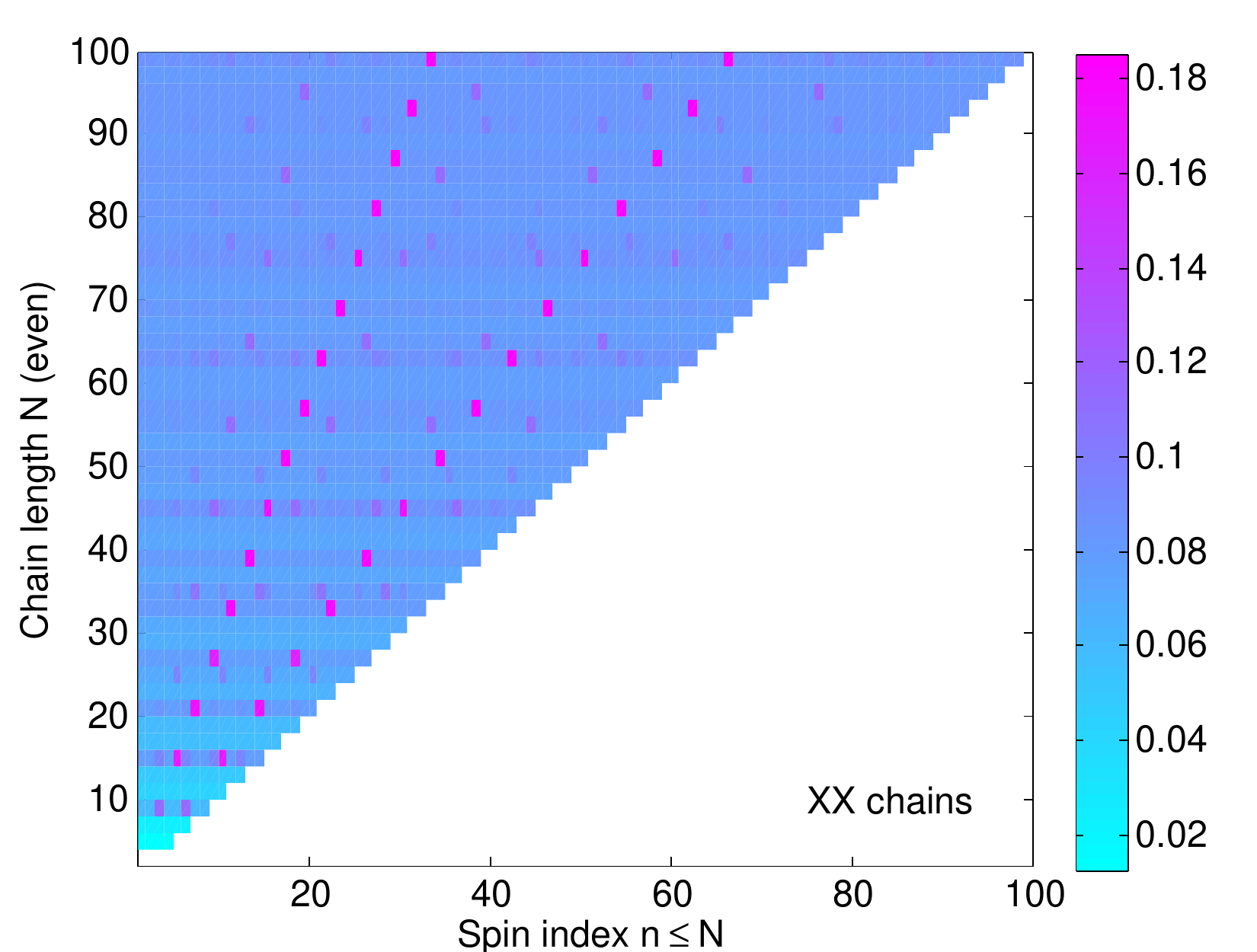}
\caption{Inertia ($\alpha=2$) versus chain length plots for XX chains.
For $N$ odd (left) the central node is a strong anti-gravity center and
weaker anti-gravity centers exist for intermediate nodes.  For $N$
(even) the anti-gravity centers are located between end and central
spins.  Again, the inertia is symmetric about the central node as
antipodal spins belong to the same equivalence class.}
\label{f:InertiaPlot-XX}
\end{figure*}

Inspired by~\cite{eurasip_clustering} we may hope that the maximum ITC
measure $d(i,j)$ defined above can be shown to be a distance.  $d(i,j)$
clearly satisfies $d(i,i)=0$ as $p(i,i)=1$, and we also have symmetry as
obviously $p(i,j)=p(j,i)$.  However, $d(i,j)$ can vanish for $i\neq j$
if $p(i,j)=1$, and in general we cannot expect the triangle inequality
to hold.  In special cases, however, such as for homogeneous chains with
either XX or Heisenberg coupling, numerical exploration for systems up
to 500 spins shows that the triangle inequality 
\begin{equation}
  D_{ijk} = d(i,k) + d(j,k)) - d(i,j) \ge 0.
\end{equation}
seems to be universally satisfied.  This renders $d(i,j)$ a
semi-distance, which induces a proper distance on a set of equivalence
classes defined by identifying nodes with $d(i,j)=0$.  Specifically, for
Heisenberg or XX chain with uniform coupling the $N$ spins can be shown
to form $\lceil\tfrac{N}{2}\rceil$ equivalence classes comprised of
spins $n$ and $N+1-n$, which we shall denote by $n$ for
$n=1,\ldots,\lceil N/2 \rceil$ in a slight abuse of notation.  On this
set of equivalence classes the ITC measure $d(i,j)$ is a metric, and it
is interesting to study the induced geometry of a spin chain with
respect to the ITC metric, and how it differs from the physical network
geometry.  Table~\ref{table1} gives the ITC geometry for Heisenberg
chains up to $N=10$, showing that it is very different from the
(trivial) physical geometry.  A Heisenberg chain of length $N=7$, for
instance, appears as a pyramid structure with an equilateral triangle as
its base formed by the vertices $a=\{1,7\}$, $b=\{2,6\}$, $c=\{3,5\}$
and $d=\{4\}$ as its apex, where $n$ refers to the index of the spin in
the chain.  Chains up to length $N=8$ can be embedded in $\RR^3$ but for
longer chains higher-dimensional spaces are required.  By Schoenberg's
theorem~\cite{Schoenberg1935} the distances $d(i,j)$ can be realized in
$\RR^d$ if and only if the Gram matrix $G=(G_{ij})$ is positive
semi-definite of rank $d$, where
\begin{equation}
  G_{ij} = \tfrac{1}{2} \left(d(i,N)^2 + d(j,N)^2 - d(i,j)^2 \right).
\end{equation}
(This is equivalent to the Cayley-Menger matrix criterion
of~\cite{Blumenthal1953}.)  We numerically verified that the Gram matrix
for both Heisenberg and XX chains is positive semi-definite for spins up
to $N=500$.  Analysis of the rank of the Gram matrix furthermore suggests
that the dimension $d$ required to embed a chain of length $N$ is
$\lceil\tfrac{N-2}{2}\rceil$ for both Heisenberg and XX chains.  Note
that Schoenberg's theorem also applies to the geometric realisability
of general spin networks and whether our ITC measure fulfills the
triangle inequality.


To better understand the geometry for very long chains we analyze its
curvature \`a la Gromov.  From both the Gromov and the scaled-Gromov
point of view~\cite{scaled_gromov} spin chain of both Heisenberg and XX
coupling type appear hyperbolic.  The strict Gromov point of view is
displayed in Fig.~\ref{f:Gromov-plot}(a) and the scaled-Gromov point of
view is shown in Fig.~\ref{f:Gromov-plot}(b).  The Gromov property might
appear to conflict with the Euclidean embeddability of the metric space
made up by the clusters with the ITC metric, as the
Bonk-Schramm~\cite{bonk_schramm} theorem says that Gromov negatively
curved spaces are embeddable in hyperbolic space.  However, some metric
spaces are embeddable in {\it both} Euclidean and hyperbolic spaces, the
most striking example being that of a complete graph with uniform edge
weight, and spin chains with uniform couplings appear to fall into this
category.

Classical communication networks, both wired~\cite{scaled_gromov,
internet_mathematics} and wireless~\cite{eurasip_clustering}, have been
shown to be Gromov hyperbolic.  Gromov hyperbolic spaces have a unique
vertex that achieves the minimum inertia~\cite{Jost1997}, the so-called
\semph{gravity center}
\begin{equation}
 g = \arg \min_{i} I(i)
   = \arg \min_{i} \sum_{j} d^\alpha(i,j), \quad \alpha \geq 1.
\end{equation}
Classical networks indeed show a point of minimum inertia, which can be
interpreted as a congestion point~\cite{IJCCC}.  Classical networks also
have the property that their Gromov boundary in the asymptotic limit is
a Cantor set~\cite{JonckheerLouHespanhaBarooahSep07}.  Quite
surprisingly, quantum networks differ from their classical counterparts
in that they have points of maximum inertia, or \semph{anti-gravity
centers}, as shown in Figs~\ref{f:InertiaPlot-H} and
\ref{f:InertiaPlot-XX} for Heisenberg and XX chains, respectively, and
their Gromov boundary is a single point.  The interpretation of the
antigravity center is that the information flow in the network avoids
this node, making it difficult to transfer excitation to and from it.
It is interesting to note in this context that the distance between
antipodal nodes is $0$, meaning that we can achieve state transfer
fidelities arbitrarily close to $1$, i.e., near perfect state transfer
between the ends of the chain of any length given sufficient time, while
near perfect state transfer is never possible between any pair of nodes
with $d(i,j)>0$, no matter how long we wait.

\section{ITC ``Topology'' Of General Networks}

The maximum transfer probability is also a useful measure for the
maximum fidelity of quantum state transfer in general spin networks.
Although it is in general no longer a metric, it can still be used as a
similarity measure to define a hierarchical clustering.  We use the
pairwise clustering algorithm introduced in~\cite{Mills2001a}.  Pairs of
nodes are grouped hierarchically into clusters in order of their
similarity, and only those clusters whose elements are closer to each
other than any element outside the cluster are preserved. If we define a
relation
\begin{equation}
  \ket{a} =_\epsilon \ket{b} :\Leftrightarrow
  d(\ket{a},\ket{b}) < \epsilon
\end{equation}
for nodes $\ket{a}$, $\ket{b}$, then these clusters are the equivalence
classes of $=_\epsilon$ for a certain $\epsilon$.  The cluster hierarchy
reveals the closeness of the nodes in terms of information transfer.

\begin{figure}%
\includegraphics[width=0.49\textwidth]{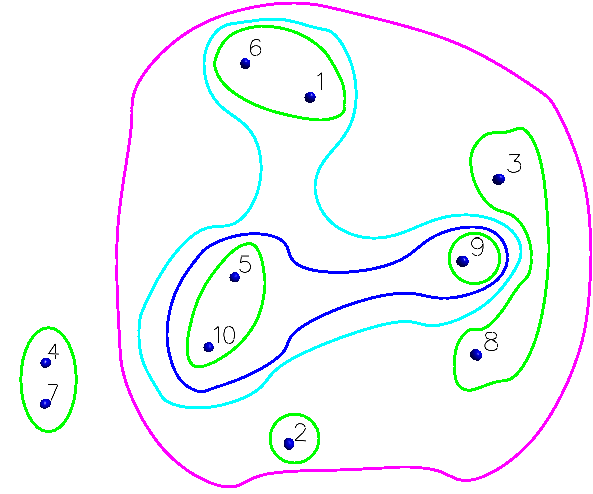}
\caption{Clustering induced by maximum transfer probability measure
for a general spin network shows that clustering differs from what
would be expected if we clustered according to physical distance.}
\label{f:Cluster-network}
\end{figure}

For example, consider a network of 10 spins distributed in a square
forming a general spin network as shown in Fig.~\ref{f:Cluster-network}.
The positions of the spins are indicated by the blue dots.  Taking the
coupling strength $J(i,j)$ between spins $i$ and $j$ to be inversely
proportional to the cube of the physical distance between the nodes, we
compute the Hamiltonian of the network, assuming XX-coupling.  We then
diagonalize this Hamiltonian and compute the maximum transfer
probabilities $p(i,j)$ and the associated $d(i,j)$, which are used as
input for our clustering algorithm.  The resulting hierarchical
clustering structure is shown in Fig.~\ref{f:Cluster-network}.
Different colours indicate clusters for different similarity levels;
i.e. for clusters marked in the same colour there exists an $\epsilon$
for which these are the equivalence classes of $=_\epsilon$.  Again, the
example shows that the physical distance of the spins is not a good
measure of their proximity in an quantum information transfer fidelity
sense.  For instance, spin $9$ is physically closer to $3$ than any of
the other nodes, yet the clustering indices that $9$ and $3$ are several
levels removed with regard to the maximum ITC.

\section{Control of Information Transfer}

To overcome intrinsic limitations on quantum state transfer or speed up
transfer, one can either try to engineer spin chains or networks with
non-uniform couplings or introduce dynamic control to change the network
topology.  The idea of engineered couplings was originally proposed to
achieve perfect state transfer between the end spin in spin chain
quantum wires~\cite{perfect-state-transfer}.  The analysis above shows
that engineering the couplings is not strictly necessary.  As the
distance between the end spins with regard to the ITC metric defined
above is zero for uniform XX and Heisenberg chains, we can achieve
arbitrary high fidelities state transfer between the end spins if we
wait long enough.  Engineering the couplings, however, can speed up
certain state transfer tasks such as state transfer between the end
spins at the expense of others.

\begin{figure}
\includegraphics[width=0.49\textwidth]{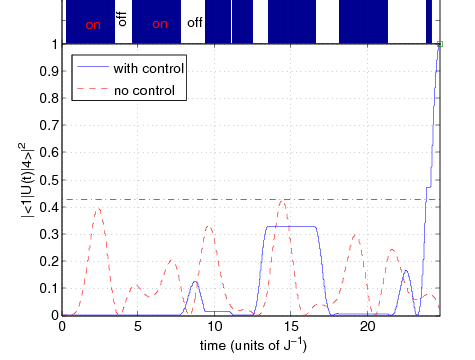}
\caption{Population of $\ket{4}$ for a uniform chain XX chain of length
$N=7$ under free and controlled evolution.  Under free evolution (dashed)
the population cannot exceed $0.4268$ (dash-dot line) but control can
overcome this restriction resulting in near perfect excitation transfer.}  
\label{f:ControlEvol}
\end{figure}

\begin{figure}
\includegraphics[width=0.49\textwidth]{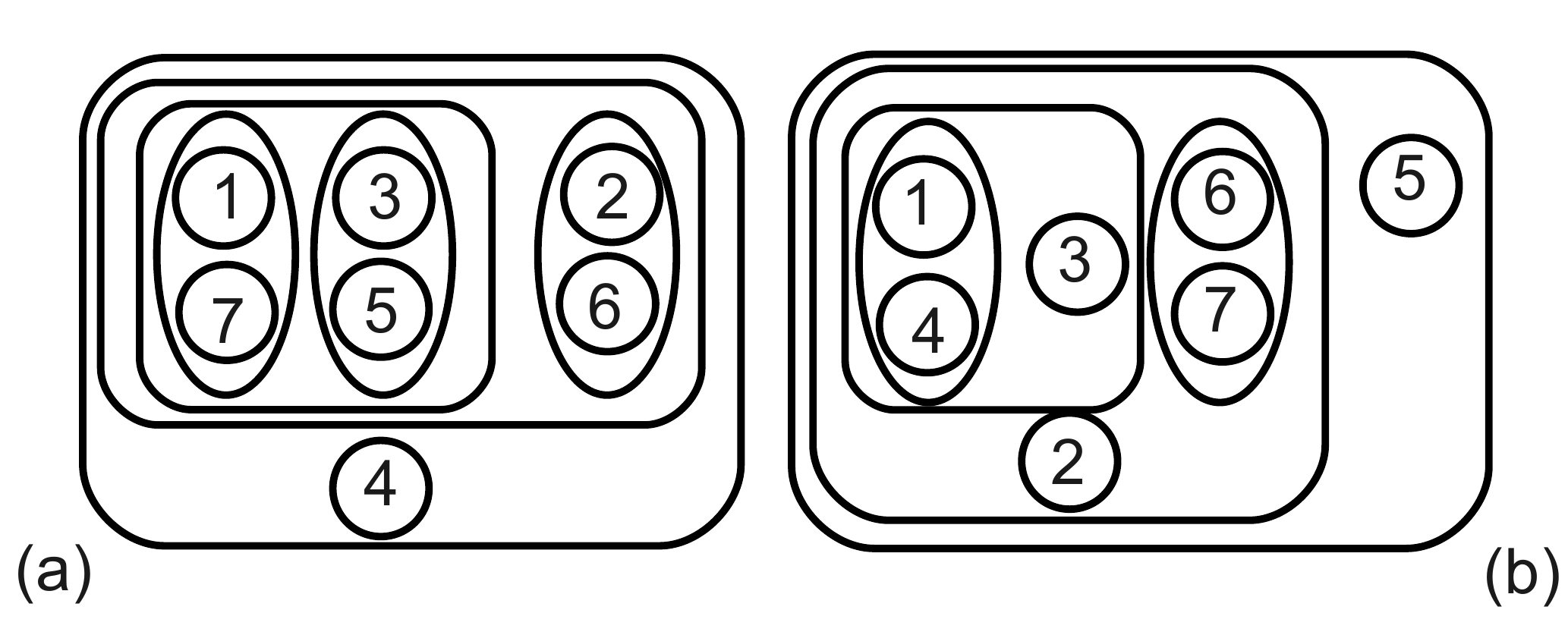}
\caption{Clustering induced by maximum transfer probability for a
uniform chain XX chain of length $N=7$ without control (a) and
clustering induced by controlled transition probability (b) for
a bang-bang control sequence designed to achieve perfect state
transfer from spin $1$ to $4$.}
\label{f:Control}
\end{figure}

A more flexible alternative to fixed engineered couplings is to apply
control.  For instance, suppose we would like to transfer an excitation
from node $1$ to $4$ for an XX chain of length $N=7$.  Node $4$ being
the anti-gravity center, the maximum transfer probability without
control is low regardless how long we are prepared to wait.  If we are
able to change the Hamiltonian of the network by applying some control
perturbtation so that $H=H_0+u(t)H_1$ we can change the situation even
if the control is restricted to $u(t)=0,1$, and $H_1$ is a local
perturbation, e.g., of a single spin induced by a magnetic field, e.g.,
$H_1=\sz^{(1)}$.  Switching the control on/off at times $t_n$ induces the
evolution
\begin{equation}
  U_{u}(t_n,0) = U_{n-1\!\!\!\mod 2}(t_n,t_{n-1}) 
                 \cdots U_1(t_2,t_1)U_{0}(t_1,t_0)
\end{equation}
where $U_0(t_k,t_{k-1})=e^{-\imath (t_k-t_{k-1})H_0}$ and
$U_1(t_k,t_{k-1})= e^{-\imath (t_k-t_{k-1})(H_0+H_1)}$.  By optimizing
the control sequence, i.e., in this restricted case the switching times
$\{t_k\}$, we can change the dynamics to achieve near perfect transfer
of an excitation or quantum state to a desired target node.  In
\cite{sophie_spin_network} it was shown that applying a simple bang-bang
control sequence to a single spin can significantly speed up quantum
state transfer between the ends of a chain, but control also allows us
to overcome fundamental limits imposed the maximum ITC, enabling us to
achieve near perfect excitation transfer to the anti-gravity center in
very short time, as shown in Fig.~\ref{f:ControlEvol}, for example.  We
can think of the control sequence as implementing an effective
Hamiltonian $H_{\rm eff}$ defined by $e^{-i H_{\rm eff} t_n} =
U_{u}(t_n,0)$ at time $t_n$.  This effective Hamiltonian differs from
the system Hamiltonian $H_0$, as do the transition probabilities.  For
comparison, we diagonalize the effective Hamiltonian, compute the
associated maximum transition probabilities $p_{\rm eff}(i,j)$, and use
hierarchical clustering to elucitate the proximity relations between
spins under the original and effective Hamiltonian.
Fig.~\ref{f:Control} shows that the results for a particular control
example.

\section{Conclusions}

Following the recent trend of {\it geometrization} of classical
communication networks, we have here developed the geometry of spin
networks using an Information Transfer Capacity metric.  Classical and
quantum networks bear the similarity that they are both Gromov
hyperbolic, with the difference that classical networks have a Cantor
Gromov boundary while spin chains have their Gromov boundary reduced to
a point.  This probably accounts for yet another discrepancy: classical
networks have a gravity center (a congestion point) while quantum
networks have an anti-gravity center, a spin difficult to communicate
with.  The broader implication of this geometrization is that it
specifies {\it where} control is necessary to overcome such limitations
of the physics.

\section{Acknowledgments}

Edmond Jonckheere acknowledges funding from US National Science
Foundation under Grant CNS-NetSE-1017881.  Sophie G Schirmer
acknowledges funding from EPSRC ARF Grant EP/D07192X/1 and Hitachi.
Frank C Langbein acknowledges funding for RIVIC One Wales National
Research Centre from WAG.

\appendix

\subsection{proof of attainability}

\begin{theorem}
\label{t:attainability} 
If the numbers $\frac{\lambda_1}{\pi},\ldots,\frac{\lambda_N}{\pi}$ are
rationally independent~\cite[Proposition 1.4.1]{KatokHasselblatt}, then
$\forall \epsilon>0$ there exists a $t>0$ large enough such that
$p(i,j)-p(i,j,t)<\epsilon$.
\end{theorem}

\begin{IEEEproof}
Set the system of units such that $\hbar=1$.  In order to reach the 
maximum probability {\it exactly}, one must find a $t$ such that
\begin{align*}
-\lambda_k t = (2m+1)\pi &\mbox{ if } \Sgn(v_{ik} v_{jk}) =-1\\
-\lambda_k t = (2m)\pi   &\mbox{ if } \Sgn(v_{ik} v_{jk}) =+1
\end{align*}
where $v_{ik}=\ip{i}{v_k}$ is the projection of the eigenstate
$\vec{v_k}$ onto the basis state $\ket{i}$, etc, and $v_{ik}^*=v_{ik}$
as the eigenstates are real for real-symmetric Hamiltonians.  In other
words, the state of the $N$-dimensional dynamical system
\begin{equation*}
  \dot{\tilde{x}}(t) = 
   -\mbox{diag}\left\{\lambda_1,...,\lambda_N\right\}, \quad \tilde{x}(0)=0 
\end{equation*}
must hit a point whose coordinates are integer multiples of $\pi$, with
the correct parity. 
As the pariety is not affected by the modulo $2\pi$ operation the
problem reduces to whether the state of the system
\begin{equation}
\dot{x}(t) = -\mbox{diag}\left\{\lambda_1,...,\lambda_N\right\} \mod
 2\pi, \quad x(0)=0 
\end{equation}
hits a point $x^*$ with coordinates $0$ or $\pi$, depending on the signs
of the various $v_{ik}v_{jk}$.  This dynamical system is the {\it linear
flow} on the $N$-torus, $\mathbb{T}^N$ and by~\cite[Proposition
1.5.1]{KatokHasselblatt} it is minimal~\cite[Definition
1.3.2]{KatokHasselblatt}, that is, the orbit of {\it every point} is
dense.  (Observe that minimality is stronger than topological
transitivity~\cite[Definition 1.3.1]{KatokHasselblatt}!)  Hence we can
get arbitrarily close to the point with desired coordinates provided we
allow $t$ to be large enough.
\end{IEEEproof}

To get a quantitative estimate of how close the state $x(t)$ has to be to
the target point $x^*$ with coordinates $0,\pi$ so that $p(i,j)-p(i,j,t) 
\leq \epsilon$ note that
\begin{align*}
\sqrt{p(i,j)} 
  =& \left| \sum_k (\pm 1) v_{ik}v_{jk} \right| \\
  =& \left| \sum_k v_{ik}v_{jk} e^{-\imath \lambda_k t} 
      + v_{ik}v_{jk}\left(\pm 1-e^{-\imath \lambda_k t} \right) \right|\\
\leq & \sqrt{p(i,j)} + \sum_k \left| v_{ij}v_{jk} 
      \left(\pm 1 -e^{-\imath \lambda_k t} \right) \right|\\
\leq & \sqrt{p(i,j)}
     + \sum_k \left| \pm 1 -e^{-\imath \lambda_k t} \right|.
\end{align*} 
Thus we have
$\sqrt{p(i,j)}-\sqrt{p(i,j,t)} \leq  
\sum_k \left| \pm 1 -e^{-\imath \lambda_k t} \right|$.
From physical considerations we know that $0\le p(i,j,t)\le 1$.
\[ 
   p_{\max}-p = \left(\sqrt{p_{\max}} -\sqrt{p} \right)
                \left(\sqrt{p_{\max}} +\sqrt{p} \right)
\]
thus shows that to secure $p_{\max}-p\leq \epsilon$, it suffices to make
$\left|\pm 1 -e^{-\imath \lambda_k t}\right|\leq \frac{\epsilon}{2N}$.
If we set $x_k(t)=-\lambda_kt$ and denote the dynamical target state as
$x^*_k:= 0 \mbox{ or } \pi$, it suffices that the target state and the
actual state are within the specification $|x^*_k-x_k(t)|\leq
\sin^{-1}\left(\frac{\epsilon}{2N}\right)$.  Since the topology induced
by hypercubes is equivalent to the usual topology induced by balls, the
latter specification can be achieved by the density of the orbit of $0$.

\subsection{Estimate of time to attain maximum probability}

The preceding material only tells us that one can reach arbitrarily
closely the maximum probability, but it does not tell us how much time
it takes.  A conservative estimate can be derived, conservative in the
sense that it assumes that the dynamical evolution in $x$ has been
discretized as the {\it translation on the
torus}~\cite[Sec. 1.4]{KatokHasselblatt},
\[ 
  x(n+1)= x(n)-\mbox{diag}\{\lambda_1,...,\lambda_N\} 
         \quad \bmod 2\pi, \quad x(0)=0
\] 
The key result is that, under the condition that
$\frac{\lambda_1}{\pi},...,\frac{\lambda_N}{\pi},1$ are rationally
independent, the {\it translation} on the torus is also
minimal~\cite[Prop. 1.4.1]{KatokHasselblatt}.  In this case, the problem
consists in finding $n \in \mathbb{N}$ such that
\begin{align*}
-\lambda_k n = (2m+1)\pi &\mbox{ if } \Sgn (v_{ik}v_{jk})=-1\\
-\lambda_k n = (2m)\pi   &\mbox{ if } \Sgn (v_{ik}v_{ij})=+1
\end{align*}
is satisfied with arbitrary accuracy, which can be derived from:

\begin{theorem}[Nowak\cite{Novak84}] 
For any $\frac{\lambda_k}{\pi} \in \mathbb{R}
\setminus \mathbb{Z}$, $k=1,...,N$, there exist infinitely many
$((p_1,...,p_N),q) \in \mathbb{Z}^N \times \mathbb{N}$ such that
\begin{equation}
  \sum_{k=1}^N \left| -\frac{\lambda_k}{\pi} - \frac{p_k}{q}\right| \leq
   \frac{c_N^{-1/N}}{q^{1+1/N}} 
\end{equation}
where the supremum, $\bar{c}_N$, of all $c_N$'s satisfying the above is
known as $\bar{c}_1=\sqrt{5}$, $\bar{c}_2=\sqrt{23}/2$, and for larger
$N$ estimated as $\bar{c}_3 \geq 1.7739$, and, for $N \geq 4$,
$\bar{c}_N \geq (N+1)^{(N+1)/2} N^{-N/2}(\pi/2)^{(N+1)/2}/\Gamma((N+5)/2)$.
\end{theorem}

We already know by the minimality of the discrete flow, except for the
error bound.  Therefore, the set of $((p_1,...,p_N),q) \in \mathbb{Z}^N
\times \mathbb{N}$ of the above theorem and the set of those that
satisfy the parity condition have a nonempty intersection.  Take a
$((p_1,...,p_N),q)$ in this intersection; thus the $p_k$ are consistent
with the parity condition.  Hence
\begin{equation*}
  \sum_{k=1}^N \left| -\frac{\lambda_k n}{\pi} - \frac{p_k}{q}n \right|
  < \frac{c_N^{-1/N}}{q^{1+1/N}} n 
\end{equation*}
Taking $n = q$ yields an $\ell^1$ error bound of $c_N^{-1/N}/q^{1/N}$ on
the $(x/\pi)$-dynamics and furthermore an error bound of $\frac{\pi
c_N^{-1/N}}{q^{1/N}} =: \epsilon$ on the $x$-dynamics.  Thus the time it
takes to be within an $\ell^1$-ball of ``radius'' $\epsilon$ around one
of the desired points is estimated as $n = q \approx \frac{\pi^N
c_N^{-1}}{\epsilon^N}$.

\end{document}